\PassOptionsToPackage{unicode}{hyperref}
\PassOptionsToPackage{hyphens}{url}
\PassOptionsToPackage{dvipsnames,svgnames,x11names}{xcolor}
\documentclass[twocolumn
]{article}
\usepackage{amsmath,amssymb}
\usepackage{lmodern}
\usepackage{iftex}
\ifPDFTeX
  \usepackage[T1]{fontenc}
  \usepackage[utf8]{inputenc}
  \usepackage{textcomp} 
\else 
  \usepackage{unicode-math}
  \defaultfontfeatures{Scale=MatchLowercase}
  \defaultfontfeatures[\rmfamily]{Ligatures=TeX,Scale=1}
\fi
\IfFileExists{upquote.sty}{\usepackage{upquote}}{}
\IfFileExists{microtype.sty}{
  \usepackage[]{microtype}
  \UseMicrotypeSet[protrusion]{basicmath} 
}{}
\makeatletter
\@ifundefined{KOMAClassName}{
  \IfFileExists{parskip.sty}{%
    \usepackage{parskip}
  }{
    \setlength{\parindent}{0pt}
    \setlength{\parskip}{6pt plus 2pt minus 1pt}}
}{
  \KOMAoptions{parskip=half}}
\makeatother
\usepackage{xcolor}
\usepackage{graphicx}
\makeatletter
\def\maxwidth{\ifdim\Gin@nat@width>\linewidth\linewidth\else\Gin@nat@width\fi}
\def\maxheight{\ifdim\Gin@nat@height>\textheight\textheight\else\Gin@nat@height\fi}
\makeatother
\setkeys{Gin}{width=\maxwidth,height=\maxheight,keepaspectratio}
\makeatletter
\def\fps@figure{htbp}
\makeatother
\NewDocumentCommand\citeproctext{}{}
\NewDocumentCommand\citeproc{mm}{%
  \begingroup\def\citeproctext{#2}\cite{#1}\endgroup}
\makeatletter
 \let\@cite@ofmt\@firstofone
 \def\@biblabel#1{}
 \def\@cite#1#2{{#1\if@tempswa , #2\fi}}
\makeatother
\newlength{\cslhangindent}
\newlength{\csllabelwidth}
\newenvironment{CSLReferences}[2] 
 {\begin{list}{}{%
  \setlength{\itemindent}{0pt}
  \setlength{\leftmargin}{0pt}
  \setlength{\parsep}{0pt}
  \ifodd #1
   \setlength{\leftmargin}{\cslhangindent}
   \setlength{\itemindent}{-1\cslhangindent}
  \fi
  \setlength{\itemsep}{#2\baselineskip}}}
 {\end{list}}
\usepackage{calc}

\ifLuaTeX
\usepackage[bidi=basic]{babel}
\else
\usepackage[bidi=default]{babel}
\fi
\babelprovide[main,import]{american}

\def\languageshorthands#1{}
\ifLuaTeX
  \usepackage{selnolig}  
\fi
\IfFileExists{bookmark.sty}{\usepackage{bookmark}}{\usepackage{hyperref}}
\IfFileExists{xurl.sty}{\usepackage{xurl}}{} 
\hypersetup{
  pdftitle={McFACTS: Monte carlo For AGN Channel Testing and
Simulation},
  pdfauthor={Jake Postiglione, Nicolas Posner, Vera Delfavero, K. E.
Saavik Ford, Barry McKernan, Harrison E. Cook, Miranda McCarthy, Emily
McPike, Kaila Nathaniel, Rosalba Perna, Varun Pritmani, Shawn Ray,
Richard O'Shaughnessy},
  pdflang={en-US},
  colorlinks=true,
  linkcolor={Maroon},
  filecolor={Maroon},
  citecolor={Blue},
  urlcolor={Blue},
  pdfcreator={LaTeX via pandoc}}

\title{McFACTS: Monte carlo For AGN Channel Testing and Simulation}

\definecolor{c53baa1}{RGB}{83,186,161}
\definecolor{c202826}{RGB}{32,40,38}
\def \rorglobalscale {0.1}
\newcommand{\rorlogo}{%
\begin{tikzpicture}[y=1cm, x=1cm, yscale=\rorglobalscale,xscale=\rorglobalscale, every node/.append style={scale=\rorglobalscale}, inner sep=0pt, outer sep=0pt]
  \begin{scope}[even odd rule,line join=round,miter limit=2.0,shift={(-0.025, 0.0216)}]
    \path[fill=c53baa1,nonzero rule,line join=round,miter limit=2.0] (1.8164, 3.012) -- (1.4954, 2.5204) -- (1.1742, 3.012) -- (1.8164, 3.012) -- cycle;
    \path[fill=c53baa1,nonzero rule,line join=round,miter limit=2.0] (3.1594, 3.012) -- (2.8385, 2.5204) -- (2.5172, 3.012) -- (3.1594, 3.012) -- cycle;
    \path[fill=c53baa1,nonzero rule,line join=round,miter limit=2.0] (1.1742, 0.0669) -- (1.4954, 0.5588) -- (1.8164, 0.0669) -- (1.1742, 0.0669) -- cycle;
    \path[fill=c53baa1,nonzero rule,line join=round,miter limit=2.0] (2.5172, 0.0669) -- (2.8385, 0.5588) -- (3.1594, 0.0669) -- (2.5172, 0.0669) -- cycle;
    \path[fill=c202826,nonzero rule,line join=round,miter limit=2.0] (3.8505, 1.4364).. controls (3.9643, 1.4576) and (4.0508, 1.5081) .. (4.1098, 1.5878).. controls (4.169, 1.6674) and (4.1984, 1.7642) .. (4.1984, 1.8777).. controls (4.1984, 1.9719) and (4.182, 2.0503) .. (4.1495, 2.1132).. controls (4.1169, 2.1762) and (4.0727, 2.2262) .. (4.0174, 2.2635).. controls (3.9621, 2.3006) and (3.8976, 2.3273) .. (3.824, 2.3432).. controls (3.7505, 2.359) and (3.6727, 2.367) .. (3.5909, 2.367) -- (2.9676, 2.367) -- (2.9676, 1.8688).. controls (2.9625, 1.8833) and (2.9572, 1.8976) .. (2.9514, 1.9119).. controls (2.9083, 2.0164) and (2.848, 2.1056) .. (2.7705, 2.1791).. controls (2.6929, 2.2527) and (2.6014, 2.3093) .. (2.495, 2.3487).. controls (2.3889, 2.3881) and (2.2728, 2.408) .. (2.1468, 2.408).. controls (2.0209, 2.408) and (1.905, 2.3881) .. (1.7986, 2.3487).. controls (1.6925, 2.3093) and (1.6007, 2.2527) .. (1.5232, 2.1791).. controls (1.4539, 2.1132) and (1.3983, 2.0346) .. (1.3565, 1.9436).. controls (1.3504, 2.009) and (1.3351, 2.0656) .. (1.3105, 2.1132).. controls (1.2779, 2.1762) and (1.2338, 2.2262) .. (1.1785, 2.2635).. controls (1.1232, 2.3006) and (1.0586, 2.3273) .. (0.985, 2.3432).. controls (0.9115, 2.359) and (0.8337, 2.367) .. (0.7519, 2.367) -- (0.1289, 2.367) -- (0.1289, 0.7562) -- (0.4837, 0.7562) -- (0.4837, 1.4002) -- (0.6588, 1.4002) -- (0.9956, 0.7562) -- (1.4211, 0.7562) -- (1.0118, 1.4364).. controls (1.1255, 1.4576) and (1.2121, 1.5081) .. (1.2711, 1.5878).. controls (1.2737, 1.5915) and (1.2761, 1.5954) .. (1.2787, 1.5991).. controls (1.2782, 1.5867) and (1.2779, 1.5743) .. (1.2779, 1.5616).. controls (1.2779, 1.4327) and (1.2996, 1.3158) .. (1.3428, 1.2113).. controls (1.3859, 1.1068) and (1.4462, 1.0176) .. (1.5237, 0.944).. controls (1.601, 0.8705) and (1.6928, 0.8139) .. (1.7992, 0.7744).. controls (1.9053, 0.735) and (2.0214, 0.7152) .. (2.1474, 0.7152).. controls (2.2733, 0.7152) and (2.3892, 0.735) .. (2.4956, 0.7744).. controls (2.6016, 0.8139) and (2.6935, 0.8705) .. (2.771, 0.944).. controls (2.8482, 1.0176) and (2.9086, 1.1068) .. (2.952, 1.2113).. controls (2.9578, 1.2253) and (2.9631, 1.2398) .. (2.9681, 1.2544) -- (2.9681, 0.7562) -- (3.3229, 0.7562) -- (3.3229, 1.4002) -- (3.4981, 1.4002) -- (3.8349, 0.7562) -- (4.2603, 0.7562) -- (3.8505, 1.4364) -- cycle(0.9628, 1.7777).. controls (0.9438, 1.7534) and (0.92, 1.7357) .. (0.8911, 1.7243).. controls (0.8623, 1.7129) and (0.83, 1.706) .. (0.7945, 1.7039).. controls (0.7588, 1.7015) and (0.7252, 1.7005) .. (0.6932, 1.7005) -- (0.4839, 1.7005) -- (0.4839, 2.0667) -- (0.716, 2.0667).. controls (0.7477, 2.0667) and (0.7805, 2.0643) .. (0.8139, 2.0598).. controls (0.8472, 2.0553) and (0.8768, 2.0466) .. (0.9025, 2.0336).. controls (0.9282, 2.0206) and (0.9496, 2.0021) .. (0.9663, 1.9778).. controls (0.9829, 1.9534) and (0.9914, 1.9209) .. (0.9914, 1.8799).. controls (0.9914, 1.8362) and (0.9819, 1.8021) .. (0.9628, 1.7777) -- cycle(2.6125, 1.3533).. controls (2.5889, 1.2904) and (2.5553, 1.2359) .. (2.5112, 1.1896).. controls (2.4672, 1.1433) and (2.4146, 1.1073) .. (2.3529, 1.0814).. controls (2.2916, 1.0554) and (2.2228, 1.0427) .. (2.1471, 1.0427).. controls (2.0712, 1.0427) and (2.0026, 1.0557) .. (1.9412, 1.0814).. controls (1.8799, 1.107) and (1.8272, 1.1433) .. (1.783, 1.1896).. controls (1.7391, 1.2359) and (1.7052, 1.2904) .. (1.6817, 1.3533).. controls (1.6581, 1.4163) and (1.6465, 1.4856) .. (1.6465, 1.5616).. controls (1.6465, 1.6359) and (1.6581, 1.705) .. (1.6817, 1.7687).. controls (1.7052, 1.8325) and (1.7388, 1.8873) .. (1.783, 1.9336).. controls (1.8269, 1.9799) and (1.8796, 2.0159) .. (1.9412, 2.0418).. controls (2.0026, 2.0675) and (2.0712, 2.0804) .. (2.1471, 2.0804).. controls (2.223, 2.0804) and (2.2916, 2.0675) .. (2.3529, 2.0418).. controls (2.4143, 2.0161) and (2.467, 1.9799) .. (2.5112, 1.9336).. controls (2.5551, 1.8873) and (2.5889, 1.8322) .. (2.6125, 1.7687).. controls (2.636, 1.705) and (2.6477, 1.6359) .. (2.6477, 1.5616).. controls (2.6477, 1.4856) and (2.636, 1.4163) .. (2.6125, 1.3533) -- cycle(3.8015, 1.7777).. controls (3.7825, 1.7534) and (3.7587, 1.7357) .. (3.7298, 1.7243).. controls (3.701, 1.7129) and (3.6687, 1.706) .. (3.6333, 1.7039).. controls (3.5975, 1.7015) and (3.5639, 1.7005) .. (3.5319, 1.7005) -- (3.3226, 1.7005) -- (3.3226, 2.0667) -- (3.5547, 2.0667).. controls (3.5864, 2.0667) and (3.6192, 2.0643) .. (3.6526, 2.0598).. controls (3.6859, 2.0553) and (3.7155, 2.0466) .. (3.7412, 2.0336).. controls (3.7669, 2.0206) and (3.7883, 2.0021) .. (3.805, 1.9778).. controls (3.8216, 1.9534) and (3.8301, 1.9209) .. (3.8301, 1.8799).. controls (3.8301, 1.8362) and (3.8206, 1.8021) .. (3.8015, 1.7777) -- cycle;
  \end{scope}
\end{tikzpicture}
}

\usepackage[affil-it]{authblk}
\usepackage{orcidlink}
\author[1,2%
  \ensuremath\mathparagraph]{Jake Postiglione%
    \,\orcidlink{0000-0003-0738-8186}\,%
    }
\author[9%
  ]{Nicolas Posner%
    \,\orcidlink{0009-0004-4600-5074}\,%
    }
\author[8%
  ]{Vera Delfavero%
    \,\orcidlink{0000-0001-7099-765X}\,%
    }
\author[1,2,3,7%
  ]{K. E. Saavik Ford%
    \,\orcidlink{0000-0002-5956-851X}\,%
    }
\author[1,2,3,7%
  ]{Barry McKernan%
    \,\orcidlink{0000-0002-9726-0508}\,%
    }
\author[4%
  ]{Harrison E. Cook%
    \,\orcidlink{0000-0001-7163-8712}\,%
    }
\author[1,2%
  ]{Miranda McCarthy%
    \,\orcidlink{0009-0005-9964-4790}\,%
    }
\author[1,2%
  ]{Emily McPike%
    \,\orcidlink{0009-0008-5622-6857}\,%
    }
\author[5%
  ]{Kaila Nathaniel%
    \,\orcidlink{0000-0003-2430-9515}\,%
    }
\author[6%
  ]{Rosalba Perna%
    \,\orcidlink{0000-0002-3635-5677}\,%
    }
\author[2%
  ]{Varun Pritmani%
    \,\orcidlink{0009-0000-3666-0586}\,%
    }
\author[1,2%
  ]{Shawn Ray%
    \,\orcidlink{0009-0005-5038-3171}\,%
    }
\author[5%
  ]{Richard O'Shaughnessy%
    \,\orcidlink{0000-0001-5832-8517}\,%
    }

\affil[1]{Graduate Center, City University of New York, 365 5th Avenue,
New York, NY 10016, USA%
    \,\protect\href{https://ror.org/00awd9g61}{\protect\rorlogo}\,%
  }
\affil[2]{Department of Astrophysics, American Museum of Natural
History, New York, NY 10024, USA%
    \,\protect\href{https://ror.org/03thb3e06}{\protect\rorlogo}\,%
  }
\affil[3]{Department of Science, BMCC, City University of New York, New
York, NY 10007, USA%
    \,\protect\href{https://ror.org/040hwr020}{\protect\rorlogo}\,%
  }
\affil[4]{New Mexico State University, Department of Astronomy, PO Box
30001 MSC 4500, Las Cruces, NM 88003, USA%
    \,\protect\href{https://ror.org/01kbvt179}{\protect\rorlogo}\,%
  }
\affil[5]{Center for Computational Relativity and Gravitation, Rochester
Institute of Technology, Rochester, New York 14623, USA%
    \,\protect\href{https://ror.org/00v4yb702}{\protect\rorlogo}\,%
  }
\affil[6]{Department of Physics and Astronomy, Stony Brook University,
Stony Brook, NY, 11794, USA%
    \,\protect\href{https://ror.org/05qghxh33}{\protect\rorlogo}\,%
  }
\affil[7]{Center for Computational Astrophysics, Flatiron Institute, 162
5th Ave, New York, NY 10010, USA%
    \,\protect\href{https://ror.org/00sekdz59}{\protect\rorlogo}\,%
  }
\affil[8]{Canadian Institute for Theoretical Astrophysics, University of
Toronto, 60 St George St, Toronto, ON M5S 3H8, Canada%
    \,\protect\href{https://ror.org/03dbr7087}{\protect\rorlogo}\,%
  }
\affil[9]{Data Science Institute, University of Chicago, 5801 S Ellis
Ave, Chicago, IL 60637, USA%
    \,\protect\href{https://ror.org/024mw5h28}{\protect\rorlogo}\,%
  }
\affil[$\mathparagraph$]{Corresponding author: %
}
\date{20 Nov 2026}

\begin{document}
\maketitle

\section{Summary}\label{summary}

\texttt{McFACTS} (Monte carlo For AGN Channel Testing and Simulation) is
the first open-source population synthesis code for modeling Active
Galactic Nuclei (AGN) (\citeproc{ref-mckernan:2025}{McKernan et al.,
2025}). AGN are gas-rich and dynamically complex structures at the
center of some galaxies. The core an AGN contain supermassive black
holes (SMBH), holding together a diverse swarm of black holes and stars
that collectively form a nuclear star cluster (NSC). A portion of the
gravitational wave (GW) events detected by the LIGO-Virgo-KAGRA (LVK)
are expected to originate from binary black holes (BBHs) embedded in AGN
disks. Some of these BBHs are expected to be detectable by the Laser
Interferometer Space Antenna (LISA) GW observatory launching in the
2030s.

Given an AGN disk model and several properties such as the SMBH and NSC
mass, \texttt{McFACTS} generates an initial population of black holes
(BH) and stars, allowing them to evolve through a variety of physical
effects. Some of the objects form binaries and eventually merge into
remnant BH which can in turn become part of another binary. Other
objects will experience chance encounters with each other throughout the
simulation. Single objects orbiting the SMBH with a mass ratio (\(q\))
of \(10^{-5}\) to \(10^{-7}\) are called Extreme Mass Ratio Inspirals
(EMRIs) and evolve through GW decay.

\texttt{McFACTS} makes use of several packages commonly used in the
scientific Python ecosystem, such as \texttt{AstroPy}
(\citeproc{ref-astropy:2022}{Astropy Collaboration, 2022}),
\texttt{NumPy} (\citeproc{ref-harris:2020}{Harris \& others, 2020}), and
\texttt{SciPy} (\citeproc{ref-scipy:2020}{Virtanen \& others, 2020}).
The AGN disk structure and properties are obtained through the
\texttt{pAGN} (\citeproc{ref-gangardt:2024}{Gangardt et al., 2024})
package, supporting both the Sirko and Goodman
(\citeproc{ref-sg:2003}{2003}). and Thompson, Quataert and Murray
(\citeproc{ref-tqm:2005}{2005}) disk models.

\section{Statement of Need}\label{statement-of-need}

Of the proposed BBH formation channels, AGN remain poorly constrained. A
population synthesis code for the AGN channel must therefore make it
possible to easily vary physical assumptions and prescriptions, while
remaining performant. The original release of \texttt{McFACTS}
demonstrated these capabilities and established the scientific value of
an open-source AGN channel code where one did not previously exist.
However, this inaugural version concentrated the simulation logic and
general bookkeeping into one main script, making it difficult to
maintain and hard for end users and contributors to modify the existing
physical assumptions. This obstacle in maintain and upgrading the code
on grew worse as the size of the \texttt{McFACTS} collaboration
increased.

The restructured version of \texttt{McFACTS}, documented in this paper,
addresses these needs directly. Utilities and code for handling physical
processes now exist as independent and reusable modular units. Instances
of new classes, representing the astrophysical objects within a
simulation, are stored and managed in a central location with duplicate
and consistency checks against the entire datastore. The input/output is
centralized and upgraded to use exchangeable handlers, each supporting
different file types and improving interoperability with other codes and
data processing pipelines. Several compute-heavy functions run as
external methods implemented in Rust, with verified data parity. The
result of all of these changes is a version of \texttt{McFACTS} that is
fast to run, simple to support, and easier to adopt by collaborators
looking to extend the functionality of \texttt{McFACTS} or integrate it
into their work. \texttt{McFACTS} is indented to be used by researchers
who model GW populations or simulate AGN disks, and observers that need
a synthetic catalog of events from the AGN channel.

Several new astrophysical features are expanded on in works by members
of the \texttt{McFACTS} collaboration and are not fully discussed in
this paper: a. Models and methods for evaluating the electromagnetic
counterparts of mergers (\citeproc{ref-mcpike:2026b}{McPike, McKernan,
et al., 2026}, in prep; \citeproc{ref-mcpike:2026}{McPike, Perna, et
al., 2026}). b. Determination of merger remnant properties via a
numerical relativity surrogate model (\citeproc{ref-ray:2026}{Ray et
al., 2026}). c.~ Application of semi-analytical gas evolution models for
binary objects (\citeproc{ref-postiglione:2026}{Postiglione et al.,
2026}, in prep).

\section{State of The Field}\label{state-of-the-field}

Several population synthesis codes targeting BBH formation and evolution
currently exist, including \texttt{COSMIC}
(\citeproc{ref-breivik:2020}{Breivik \& others, 2020}), \texttt{COMPAS}
(\citeproc{ref-riley:2022}{Riley \& others, 2022}), \texttt{CMC}
(\citeproc{ref-rodriguez:2022}{Rodriguez \& others, 2022}), and
\texttt{cogsworth} (\citeproc{ref-wagg:2024}{Wagg \& others, 2024}).
These codes focus on formation channels for BBHs in gas-poor
environments (e.g.~isolated binaries, dense stellar clusters, etc.), but
do not model the critical gas-rich dynamics of an AGN disk. Codes that
focused on the AGN channel have historically been geared towards
specific features, such as focusing on the mergers found at migration
traps, or are not open source and have not been released publicly.
\texttt{McFACTS} distinguishes itself from its predecessors by being
public, open-source, and tooled to simulate the AGN channel for BBH
mergers across a broad range of the AGN parameter space.

\section{Software Design}\label{software-design}

\begin{figure}
\centering
\includegraphics[width=\linewidth,height=2.5in,keepaspectratio]{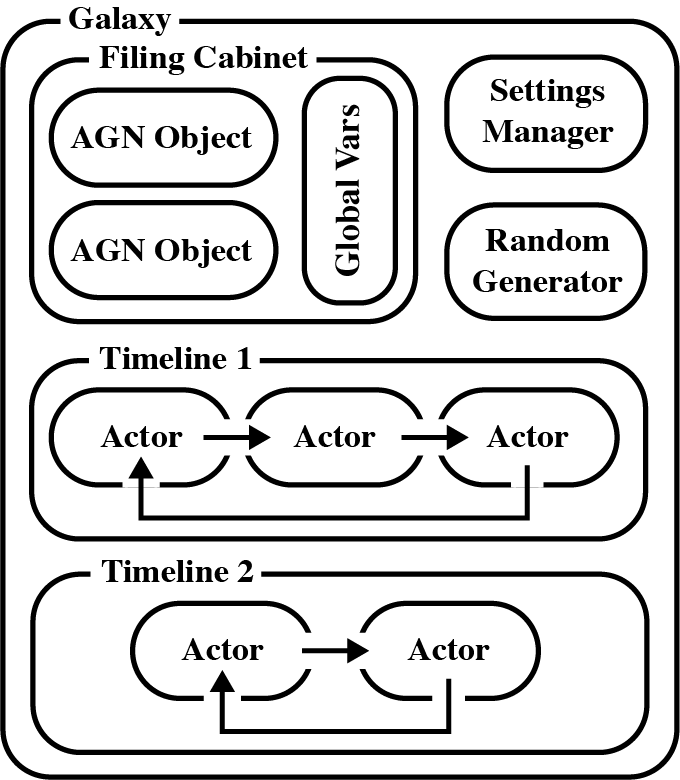}
\caption{Diagram showing a high level overview of the relationship
between different object types used by the restructured version of
\texttt{McFACTS}.}\label{diag}
\end{figure}

The restructure of \texttt{McFACTS} separates the framework from any
particular simulation or model assembled out of its building blocks.
This new framework is best understood through a top-down lens, where
each level orchestrates the level beneath it. A \texttt{Galaxy} holds
its populations in a \texttt{FilingCabinet} of \texttt{AGNObjectArray}s.
The \texttt{AGNObjectArray}s are evolved according to a
\texttt{SimulationTimeline}, and each timeline is composed of individual
\texttt{TimelineActor}s (graphically represented in \autoref{diag}). The
population is seeded with \texttt{GalaxyPopulator} objects.

\subsection{Galaxy}\label{galaxy}

The \texttt{Galaxy} object sits at the core of the simulation framework.
A single \texttt{Galaxy} object represents one synthetic galaxy, with
its own population, deterministic random state using a Philox
pseudo-random number generator, and the history of simulation timelines
that have been run. A \texttt{Galaxy} is first populated using
\texttt{GalaxyPopulator} objects. Once the \texttt{Galaxy} is seeded, a
\texttt{SimulationTimeline} holding several \texttt{TimelineActor}
objects is run.

\subsection{Filing Cabinet and AGN Object
Array}\label{filing-cabinet-and-agn-object-array}

A \texttt{FilingCabinet} is a flexible container that stores the
population for a single \texttt{Galaxy} object or an entire universe of
galaxies. The cabinet stores a dictionary of \texttt{AGNObjectArray}s
and provides a storage interface for global variables that need to be
accessed by independent modules. Consistency checks ensure an object
with the same \texttt{UUID} does not appear in multiple arrays stored in
the \texttt{FilingCabinet}. The \texttt{AGNObjectArray} object allows
for the association of attributes representing physical properties of
objects, while maintaining the vectorization convention of storing
object properties (orbital parameters, mass, spin and spin angle, etc.)
in parallel \texttt{NumPy} arrays. Subclasses for different population
categories, such as single black holes, BBHs, merged black holes, and so
forth, allow for quick object-type validation, and the inheritance of
shared attributes.

\subsection{Simulation Timeline, Timeline Actors, and Galaxy
Populators}\label{simulation-timeline-timeline-actors-and-galaxy-populators}

Evolution of the population in a \texttt{Galaxy} is scheduled through a
\texttt{SimulationTimeline} object. The timeline holds an ordered list
of different operations that are applied to the population over some
fixed number of timesteps. A single \texttt{Galaxy} can run several
timelines sequentially, processing the population and other parameters
in distinct stages. A \texttt{TimelineActor} object encapsulates the
implementation of physical processes (migration, binary formation, etc.)
through a common \texttt{perform} interface. Each actor is passed a
reference to the \texttt{FilingCabinet} of that galaxy, and can read and
write directly to it. A \texttt{GalaxyPopulator} is structured similarly
to a \texttt{TimelineActor}, but distinctly plays the role of seeding
the initial populations, constructing the starting
\texttt{AGNObjectArray} objects passed to the \texttt{FilingCabinet}.

\subsection{Scaling with SMBH Mass}\label{scaling-with-smbh-mass}

The primary executable in the restructured workflow supports scaling
several properties of the AGN with changes to input parameters. When
enabled, scaling can adjust the NSC mass, the outer disk radius, the
boundary between the inner/outer disk, the type-I migration trap radius,
and capture rates from the NSC, following the scaling relations of
Delfavero et al. (\citeproc{ref-delfavero:2025}{2025}). Prior to the
restructure, these parameters needed to be tuned by hand or modified by
a script using regular expressions. Because this step happens during
initialization, the settings written and logged in the output directory
are the final scaled settings used in the run. As a result, it is easier
to simulate AGN of different types and sizes.

\subsection{Crosscutting Utilities}\label{crosscutting-utilities}

The \texttt{SettingsManager} acts as a central repository for built-in
defaults and user-overridable settings. Internal constants are protected
from modification by the user, and custom settings can be passed for use
with third-party modules. The \texttt{SnapshotHandler} provides an
interface for the simulation framework to load and save
\texttt{SettingsManager} and \texttt{FilingCabinet} objects. By default,
we support both \texttt{.txt} and \texttt{.ini} for the loading and
saving of the \texttt{SettingsManager} and support \texttt{.txt} files
for the \texttt{FilingCabinet}. Members of the \texttt{McFACTS}
collaboration have developed tooling to support \texttt{.hdf5} files.
There are also plans to support \texttt{.parquet} files using the Apache
Parquet format and \texttt{.db}, \texttt{.sqlite} files through SQLite.
IO formatting is highly customizable, allowing for \texttt{McFACTS} to
be easily implemented into third-party data processing pipelines and
software stacks.

\subsection{Extension: mcfast}\label{extension-mcfast}

\texttt{mcfast} is a custom-built extension to the \texttt{McFACTS}
codebase written in Rust, with Python bindings produced by
\texttt{pyo3}. It provides optimized variants of several
\texttt{McFACTS} functions, including vectorized, single-pass variants
of computationally heavy functions (e.g.~\texttt{shock\_luminosity},
\texttt{analytical\_kick\_velocity}) and tailored variants of unit
conversion functions (e.g.~\texttt{si\_from\_r\_g},
\texttt{r\_g\_from\_units}) which bypass AstroPy's unit conversion
operations.

\subsubsection{Motivation, Design, and
Parity}\label{motivation-design-and-parity}

The most time-intensive operations included long chains of
\texttt{NumPy} array operations on several arrays of equal length, and
frequently-called helper functions reliant on \texttt{AstroPy} unit
conversion operations. Several of the \texttt{mcfast} function variants
have measured a 10x to 200x performance increase over their Python
counterparts. Overall, these individual speedups have contributed to an
overall \textasciitilde6x improvement in the total runtime of the
simulation.

The primary goal of \texttt{mcfast} is to minimize data transfer across
the Python-Rust Foreign Function Interface (FFI). The inputs and outputs
of candidate functions in \texttt{McFACTS} are \texttt{NumPy} arrays, so
\texttt{mcfast} makes heavy use of zero-copy reads using the
\texttt{NumPy} crate API. Output \texttt{NumPy} arrays are allocated
from inside the Rust extension, passing only pointers across the FFI
without copying data.

\texttt{mcfast} variant functions are tested against the original Python
functions in \texttt{McFACTS}, ensuring that both functions produce the
same result. Where an exact match can't be guaranteed between the
results, a deviation of at most 1e-6 is enforced for as a tolerance for
floating point operations.

\section{Research Impact Statement}\label{research-impact-statement}

Since its initial release, \texttt{McFACTS} has been used as the primary
research tool in several publications aimed at studying the AGN channel.
As of November 2026, these publications have collectively been cited
over 100 times, and provided a wealth of insight into how the AGN
channel contributes to the population of detectable sources for LVK and
LISA. Papers published by the \texttt{McFACTS} collaboration study the
effects of AGN on the \(q\)--\(\chi_{\rm eff}\) distribution of LVK
detectable BBH mergers (\citeproc{ref-cook:2025}{Cook et al., 2025}),
show the effects of NSC properties, galaxy mass, and redshift across an
entire population of galaxies (\citeproc{ref-delfavero:2025}{Delfavero
et al., 2025}), demonstrate the effects of the AGN channel on
electromagnetic signatures and remnant kicks of embedded BBH mergers
(\citeproc{ref-mcpike:2026}{McPike, Perna, et al., 2026}), and explore
the AGN channel's effects on the resulting properties of BBH merger
remnants (\citeproc{ref-ray:2026}{Ray et al., 2026}) through numerical
relativity surrogate models.

\section{AI Usage Disclosure}\label{ai-usage-disclosure}

The authors affirm that generative artificial intelligence (AI) was not
used to produce this manuscript or the code contributions detailed by
this manuscript.

We respectfully request that the content of this manuscript is not used
to train generative models.

\section{Acknowledgements}\label{acknowledgements}

Support for JP was provided by Schmidt Sciences, LLC. JP, NP, KESF, BM,
EM, SR and MM would like to thank the Center for Computational
Astrophysics at the Flatiron Institute for their hospitality while a
portion of this research was carried out. The Flatiron Institute is a
division of the Simons Foundation. JP would like to thank Michael
Zingale for their helpful presentation and conversation about code
structuring and numerical methods.

\section*{References}\label{references}
\addcontentsline{toc}{section}{References}

\protect\phantomsection\label{refs}

\end{document}